\documentclass[10pt,twocolumn,twoside]{IEEEtran} 
   \usepackage[pdftex]{graphicx}
  \DeclareGraphicsExtensions{.pdf,.jpeg,.png}
\usepackage[]{units}
\usepackage[T1]{fontenc} 
\usepackage{amsmath}
\usepackage{amssymb}
\usepackage{bm} 
\usepackage{eurosym}
\usepackage[super]{nth} 
\usepackage{multirow}
\usepackage{color}
\usepackage[caption=false,font=footnotesize]{subfig}
\usepackage{enumitem}
\usepackage{cite}

\usepackage{comment}
\usepackage{float}

\usepackage{algorithm,caption}
\usepackage{algpseudocode} 
\usepackage{multicol}
\usepackage{tabularx}
\usepackage{eurosym}
\renewcommand{\baselinestretch}{0.943}

\begin{document}
%
\title{Decentralized Model-free Loss Minimization in Distribution Grids with the Use of Inverters}
%
%
%

\author{Ilgiz~Murzakhanov,~\IEEEmembership{Student~Member,~IEEE,}
        Spyros~Chatzivasileiadis,~\IEEEmembership{Senior~Member,~IEEE}
\thanks{This work is supported by the ID-EDGe project, funded by Innovation Fund Denmark, Grant Agreement No. 8127-00017B, and by the FLEXGRID project, funded by the European Commission Horizon 2020 program, Grant Agreement No. 863876.}
\thanks{I. Murzakhanov and S. Chatzivasileiadis are with the Department of Wind and Energy Systems, Technical University of Denmark (DTU), Kgs. Lyngby, Denmark. E-mail: \{ilgmu, spchatz\}  @dtu.dk.}}

%
%

\markboth{}%
{Murzakhanov and Chatzivasileiadis: Decentralized model-free loss minimization in distribution grids with the use of inverters}
%



\maketitle

\begin{abstract}
Distribution grids are experiencing a massive penetration of fluctuating distributed energy resources (DERs). As a result, the real-time efficient and secure operation of distribution grids becomes a paramount problem. While installing smart sensors and enhancing communication infrastructure improves grid observability, it is computationally impossible for the distribution system operator (DSO) to optimize setpoints of millions of DER units. This paper proposes communication-free and model-free algorithms that can actively control converter-connected devices, and can operate either as stand-alone or in combination with centralized optimization algorithms. We address the problem of loss minimization in distribution grids, and we analytically prove that our proposed algorithms reduce the total grid losses without any prior information about the network, requiring no communication, and based only on local measurements. Going a step further, we combine our proposed local algorithms with a central optimization of a very limited number of converters. The hybrid approaches we propose have much lower communication and computation requirement than traditional methods, while they also provide performance guarantees in case of communication failure. We demonstrate our algorithms in four networks of varying sizes: a 5-bus network, an IEEE 141-bus system, a real Danish distribution system, and a meshed IEEE 30-bus system.
\end{abstract}


%
\IEEEpeerreviewmaketitle

\section{Introduction}
%
%
%
%

\IEEEPARstart{M}{odern} distribution grids are characterized by the rapidly increasing penetration of DERs, especially photovoltaics (PVs) and battery storage systems. Reverse power flows and a greater ratio of fluctuating generation at the local level require the real-time efficient and secure operation of distribution grids. Considering the millions of DER units to be connected to the grid, however, it is almost impossible to manage their operation centrally in real-time. The computation and communication requirements for such a task go beyond the current capabilities of state-of-the art computation and communication infrastructure. Even if distributed algorithms are employed, it is improbable to have established a communication channel with all devices at all times. Parts of the grid will probably remain unobservable, or data will not be able to transmitted in real time. Therefore, communication-free (local) and model-free algorithms, which do not require any knowledge of the surrounding system are expected to play a significant role in the managing of such a system. Such algorithms, being agnostic to the topology of the system or the point where the device is connected, do not only offer plug'n'play capabilities, but, if designed appropriately, they can achieve system-wide objectives (e.g. optimal voltage profile, minimum losses, etc.) with local actions.

In this paper, we focus on the loss minimization problem and propose solutions that attempt to combine the best of both worlds. We design communication-free and model-free algorithms which \emph{provably} reduce the system losses without any prior information about the network, requiring no communication, and based only on local measurements. The analytical proofs for the performance of the two algorithms are included in \cite{arxiv_external}. Going a step further, we combine these algorithms with a central optimization of a very limited set of resources. Centralized optimization algorithms can arrive at the global optimum and can solve efficiently in real-time if the number of centrally-controlled resources remains low. Combining local algorithms with central optimization has much lower requirements for communication and computation than traditional methods, while it also provides performance guarantees in the case of communication failure.

The problem we are addressing in this paper is the minimization of distribution grid losses through the reactive power control of power electronic inverters. Despite the wide deployment of DERs, which can reduce power flows, power losses remain one of the main problems of distribution grids: electricity losses from the power plant to the consumer are around 6\% in Denmark and at least 19\% in India \cite{databank}. Grid losses, however, can substantially reduce through the active control of converter-connected devices. In the rest of this paper, we focus on the control of solar PV inverters, but our approaches can apply for any type of converter-connected device. Willing to avoid any direct control of the active power setpoint, as PV inverters, batteries, electric vehicles, and others, pay or are getting paid based on the active power they consume or inject, our algorithms only adjust the reactive power injection of the solar PV inverters within the permissible limits; this is limited by the maximum apparent power of the inverter and current active power generation of the PV panel \cite{c1}. As a matter of fact, as revealed in \cite{Goncalves2019}, a smart inverter allowing a variable power factor can achieve the lowest power losses in the grid.  

\subsection{Literature Review}
Control of PV inverters may have several goals, including minimization of active power losses and improvement of voltage profile. All approaches explored during the literature review, can be classified by the presence or absence of a central coordinator. Two approaches of controlling PV inverters' settings in a fully centralized manner have been proposed in \cite{Prasanna2015} and \cite{Horowitz2018a}. In \cite{Prasanna2015}, the authors suggest minimizing active power losses by reconfiguring network topology, while in \cite{Horowitz2018a}, centralized dynamic programming and approximate dynamic programming are utilized for decreasing voltage violations. The combined central and local control schemes for power loss minimization have been introduced in \cite{Chistyakov2012a} and \cite{Yeh2012a}. While in both works, the local control utilizes the local measurements for fast reaction during changes of load and generation, and the centralized control provides optimization of the local control units, \cite{Yeh2012a} additionally presents a linearized power flow model. A combined control scheme for improving voltage profile has been applied in \cite{Bidgoli2018a}. The local control utilizes piecewise linear $VQ$ characteristics, and the centralized controller uses model predictive control (MPC) to bring voltages inside tighter limits.
 
Next, we consider methods without a central coordinator, as it is the main focus of our work. All these methods can be evaluated according to their need for communication between agents. The reduction of power losses by a decentralized chance-constrained control policy for inverters has been proposed in \cite{Hassan2018a}. While the presented results indicate the effectiveness of the approach, communication between neighbor nodes is required. In \cite{Xiao2017a}, for optimal power control in distribution networks, the agents compute their weight matrix and exchange it with others. Several works apply the alternating direction method of multipliers (ADMM) for optimal control of distributed generation for power loss minimization. For example, the realization of the semidefinite programming (SDP) with the use of ADMM has been presented in \cite{DallAnese2013a}. The model-free decentralized algorithm for minimizing power losses has been proposed in \cite{Ahn2013}. The performance of the introduced two-level algorithm appears to be highly dependent on the communication network: the version without communication does not reach the minimum loss condition, has slower convergence and fluctuating performance. In \cite{Bolognani2015}, the distributed reactive power control algorithm for loss minimization is designed. In addition to information exchange between the neighbor nodes, the algorithm needs information about the voltage angles, which is not commonly measured in distribution grids. In \cite{saverio_arxiv}, the authors propose an Online Feedback Optimization (OFO) model-free approach, that tracks the solution of the AC-OPF under time-varying conditions. The proposed tool can learn the model sensitivity, which eliminates the requirement of having an accurate grid model and full grid observability. However, the conducted sensitivity estimation and convergence analysis in \cite{saverio_arxiv} cover the objective of only penalizing voltage deviations but not power loss minimization. In addition, the OFO still requires communication between various units.

Finally, we discuss works without a central coordinator and requiring no communication, however, all these approaches need some information about the system. In \cite{Mousa2019}, the authors have presented an affinely adjustable robust counterpart (AARC) approach for improving voltage profile. The approach requires information about line parameters. A decentralized impedance-based adaptive droop method for power loss reduction has been presented in \cite{Oureilidis2016a}. As the droop coefficients depend on the microgrid impedance, information on the electrical parameter of the connection lines is needed. Another work \cite{Ghosh2014} develops a droop algorithm for voltage control by reactive power injections from PV inverters, but the proposed droop control is based on heuristic rules. As a result, there is no guarantee of proper work of the algorithm for the power system, which topology is unknown. In \cite{Kundu2013}, the local strategy algorithm uses a parameter that is computed as the reactance to the resistance ratio of distribution lines. Similarly, the control of distributed PV generators in \cite{Jabr2018} exploits information of the network nodal admittance matrix. In \cite{Weckx2016}, the authors propose designing an optimal $Q(P)$ curve that keeps the voltage within the limits. The drawback of the proposed approach is the requirement for extensive voltage and PV output data. Additionally, the method can arrive at the state with higher power losses compared to the scenario without reactive power control. In contrast to all considered methods, our solution has a proven mathematical guarantee on minimization of power losses for any radial system without a need for any non-local information.

\subsection{Main Contributions}
The contributions of this work are the following:
\begin{itemize}
    \item We propose two model-free and communication-free algorithms, which do not need any information about the system and do not require information exchange. For any radial system, we prove that our algorithms provide lower active power losses than the ``no-action'' strategy, i.e., a scenario without reactive power control.
    \item We analytically prove that if inverters have such reactive capacity, they should be set to a value higher than the load connected to the same bus.
    \item We propose two hybrid algorithms, which incorporate proposed model-free and communication-free algorithms and further enhance performance results with the use of a central coordinator. The proposed hybrid algorithms achieve the same loss minimization results as optimal power flow (OPF) but have a much lower computation and communication burden.
    \item We validate our analytical derivations on an IEEE 141-bus radial system, a real Danish distribution system, and a meshed IEEE 30-bus system. For the 141-bus radial system, we model different numbers and locations of PVs. Moreover, we model the topology changes of the 141-bus radial system, which may occur due to the fault of a line or scheduled maintenance. For the Danish distribution system, we validate our algorithms under varying consumption and active power generation using full-year data from 2019. We demonstrate applicability of our algorithms for meshed systems and systems with various equipment (switched capacitors, transformers, load tap changers) on example of the meshed 30-bus system.
\end{itemize}

\subsection{Outline}
The remainder of this paper is organized as follows. First, the problem formulation is given in Section~\ref{sec:problem_form}. In Section~\ref{sec:inverter}, we introduce the realistic operational limits of PV inverters. Section~\ref{sec:DistFlow} contains the description of the utilized DistFlow model.
Section~\ref{sec:PropSol} presents our proposed algorithms. The numerical results are provided in Section~\ref{sec:NumRes}. Finally, Section~\ref{sec:conclus} concludes the paper and proposes future directions.

\section{Problem formulation} \label{sec:problem_form}
In this section, we introduce terms, assumptions, and variables, which are used throughout the paper. By common terminology in power systems, we use the terms buses and nodes, and branches and lines interchangeably. 

\begin{figure}[H]
    \centering
    \includegraphics[width=0.9\linewidth]{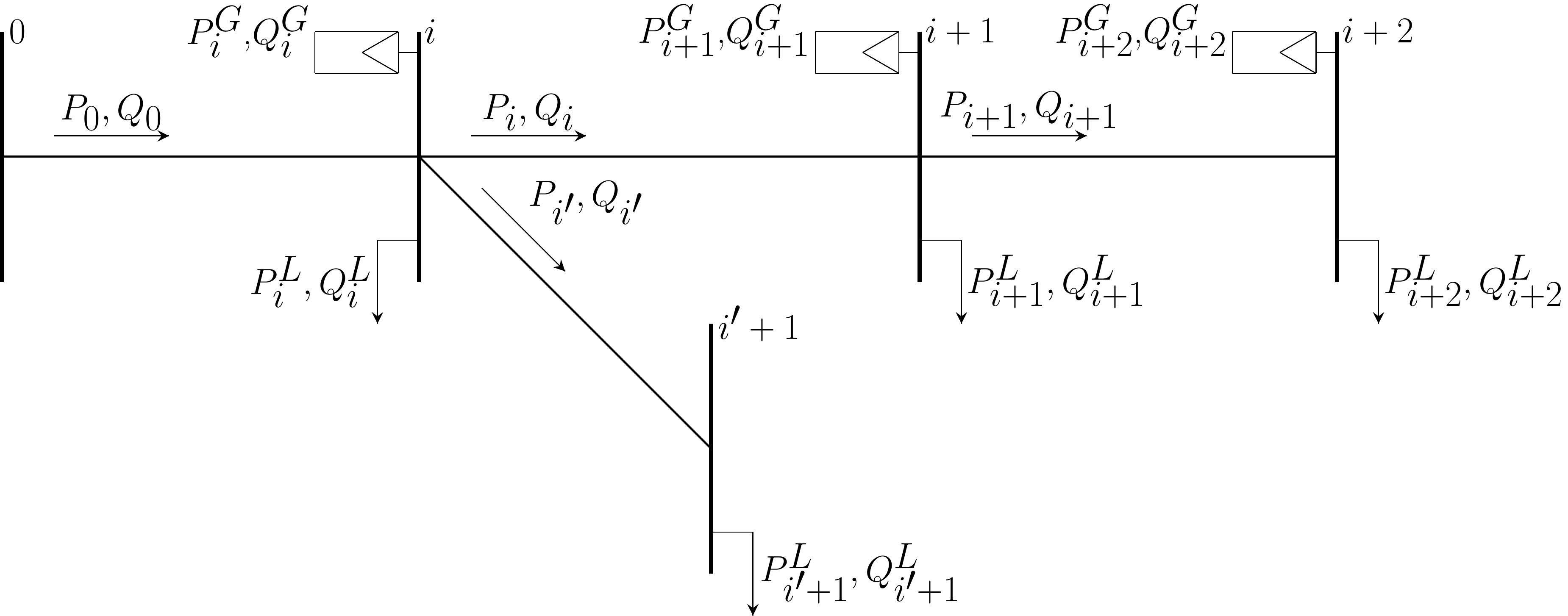}
    \caption{The 5-bus power system.}
    \label{fig:Math_System}
\end{figure}

To present our algorithms and demonstrate their performance, we use a representative 5-bus network, as shown in Fig.~\ref{fig:Math_System}. This 5-bus system is simple but sufficient to describe the concept of the solution we propose.
In Fig.~\ref{fig:Math_System}, bus $0$ is a \textit{slack bus}: it is assumed to be connected to the external grid or is, at least, able to inject (or withdraw) sufficiently large amounts of active and reactive power. 
For the sake of generality, we consider that not all nodes in the 5-bus system have a PV panel and an inverter. By that, nodes $i$, $i+1$, $i+2$ have both a load and a PV panel with an inverter, while node $i'+1$ has only a load.

The grid buses are linked through four distribution lines. By common terminology, nodes $i'+1$ and $i+2$ are called \textit{leaf nodes}. Nodes $i$ and $i+1$ are called \textit{branch nodes}, as they are placed between the slack and leaf nodes.

The branch $i$ is called a downstream branch for node $i$. As it can be seen in Fig.~\ref{fig:Math_System}, leaf nodes $i'+1$ and $i+2$ do not have downstream branches. Branch $0$ is called an upstream branch for node $i$. 

\section{Reactive power capability of PV inverter}\label{sec:inverter}
Before we introduce the algorithms for reactive power dispatch, we first discuss the limitations of the PV inverters' reactive power capability. We adopt a model of real PV inverters, with the following characteristics. 

First, the inverters' rated apparent power is equal to the rated active power \cite{InverterSpec}:
\begin{equation}\label{eq:SeqP}
\overline{S} = \overline{P}^G
\end{equation}

Second, the inverters can control their power factor from 0.8 over-excited to 0.8 under-excited \cite{InverterSpec}. Deriving the corresponding maximum angle $\phi^{max}$ and utilizing the relation between $cos$ and $tan$, we can express these limits via active and reactive power generation:
\begin{equation}\label{eq:PF}
-tan(\phi^{max}) \leq \frac{Q^G}{P^G} \leq tan(\phi^{max})
\end{equation}

Third, at each moment of time, the apparent power constraint should be satisfied:
\begin{equation}\label{eq:Slim}
|Q^G| \leq \sqrt{\overline{S}^2 - (P^G)^2}
\end{equation}

\begin{figure}[H]
    \centering
    \includegraphics[width=0.6\linewidth]{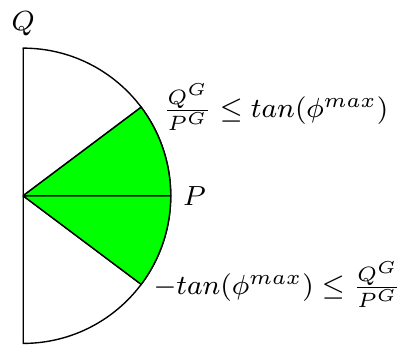}
    \caption{Phasor diagram of PV inverter.}
    \label{fig:PQDiag}
\end{figure}

The constraints (\ref{eq:SeqP})-(\ref{eq:Slim}) are described by the phasor diagram in Fig.~\ref{fig:PQDiag}. 

On a clear day with the sun angle aligned with the PV array, solar panels produce their rated capacity $P^G = \overline{P}^G$. Then, according to (\ref{eq:SeqP}) and (\ref{eq:Slim}), $Q^G = 0$. However, PV output $P^G$ is time-varying, and most of the times $P^G < \overline{P}^G$ due to clouds and the sun position in the sky. Consequently, the range of reactive power capability varies throughout the year, the day, and the weather. To model this effect, we perform our simulations for different values of $P^G$ while keeping the apparent power capability $\overline{S}$ fixed.

\section{DistFlow model}\label{sec:DistFlow}
The power flow in a radial distribution network can be described by a set of recursive equations, called \textit{DistFlow branch equations} \cite{DistFlow}. To illustrate them, consider the radial network in Fig.~\ref{fig:Math_System}. 

We represent the lines with impedances $z_l = r_l + jx_l$, the apparent power demand as $S^L = P^L + jQ^L$ and the generation as $S^G = P^G + jQ^G$. 

The equations use the active power, reactive power, and voltage magnitude at the sending end of a branch, $P_i$, $Q_i$, $V_i$ respectively to express the same quantities at the receiving end of the branch as follows.

\begin{subequations}
\begin{alignat}{2}
& P_{i+1} = P_i - r_i\frac{P^2_i + Q^2_i}{V_i^2} - P^L_{i+1} + P^G_{i+1} \label{DistFlow_P} \\
& Q_{i+1} = Q_i - x_i\frac{P^2_i + Q^2_i}{V_i^2} - Q^L_{i+1} + Q^G_{i+1} \label{DistFlow_Q} \\
& V^2_{i+1} = V^2_i - 2(r_iP_i + x_iQ_i) + (r^2_i+x^2_i)\frac{P^2_i + Q^2_i}{V_i^2} \label{DistFlow_V}
\end{alignat}
\end{subequations}

For loss reduction, the objective is to minimize the total $i^2r$ losses in the system \cite{DistFlow}; so, the power loss $\Delta P$ is defined as in (\ref{DistFlow_los}):
\begin{equation}\label{DistFlow_los}
\Delta P = \sum_{i=0}^{n-1} r_i\frac{P^2_i+Q^2_i}{V^2_i}~~~p.u.
\end{equation}
Applying (\ref{DistFlow_los}) to the 5-bus system in Fig.~\ref{fig:Math_System} results in:
\begin{equation}\label{Totlos_5bus_Gen}
\begin{split}
& \Delta P^{\mathcal{A}} = r_0\frac{P^2_0 + (Q_0^\mathcal{A})^2}{V^2_0} + r_i\frac{P^2_i+(Q_i^\mathcal{A})^2}{(V_i^\mathcal{A})^2} \\
& + r_{i'}\frac{P^2_{i'}+(Q_{i'}^\mathcal{A})^2}{(V_{i'}^\mathcal{A})^2} + r_{i+1}\frac{P^2_{i+1}+(Q_{i+1}^\mathcal{A})^2}{(V_{i+1}^\mathcal{A})^2}
\end{split}
\end{equation}
where $\mathcal{A}$ refers to a utilized algorithm. For compactness, we use the following notations for the algorithms: $\mathcal{N}$ -  the ``no-action'' strategy, ${\mathcal{H}}$ - the local load measuring algorithm, $\mathcal{F}$ - the local flow measuring algorithm. Description of the algorithms is given in Section \ref{sec:PropSol}. As it can be seen from (\ref{Totlos_5bus_Gen}), active power losses for all the algorithms consist of four terms, as there are four power lines. Note that the resistance of lines $r$, PV outputs $P$, and the voltage magnitude of the slack bus $V_0$ are kept the same for a fair comparison across the loss minimization algorithms and, therefore, the superscript specifying the algorithm for them is omitted.

\section{Proposed solution}\label{sec:PropSol}
In this section, we propose two algorithms that do not require any communication. The proposed algorithms work for \emph{any} power distribution system and both require only local information for their execution. As a result, we do not need to have knowledge of (or assume) the number or the location of such inverters in the system, as our algorithms are built communication-free and model-free, requiring no non-local information.  

To reap the benefits of both centralized and local approaches, in the third part of this section we further propose two hybrid algorithms, where we show how the two communication-free and model-free approaches can be best combined with centralized optimization algorithms that communicate setpoints only to a limited number of devices. We explore the performance of all the approaches we proposed in the Section~\ref{sec:NumRes}, where we discuss about numerical results.

\subsection{Local Load Measuring Algorithm (LLMA)}
This algorithm is inspired by the heuristic approach first proposed in \cite{c1}. We further extend this approach and provide mathematical guarantees about its performance. We name our solution the local load measuring algorithm (LLMA) and denote the corresponding variables with superscript ${\mathcal{H}}$.  For each inverter following LLMA, the only needed information is the reactive power load at the same node. We denote the reactive power limits of the inverter, which satisfy the constraints (\ref{eq:PF})-(\ref{eq:Slim}), by $\overline{Q}^G$.
\begin{algorithm}
\caption*{\textbf{Algorithm 1: Local Load Measuring Algorithm (LLMA)}} \label{alg:LLM}
\begin{algorithmic}
    \If {$\overline{Q}^G\geq Q^L$}
        \State $Q^{G,{\mathcal{H}}} = Q^L$
    \Else
        \State $Q^{G,{\mathcal{H}}} = \overline{Q}^G$
    \EndIf
\end{algorithmic} 
\end{algorithm}

Algorithm 1 also has a closed equivalent form, which includes all the constraints explicitly:
\begin{equation}\label{eq:LLMA}
Q^{G,{\mathcal{H}}} = \min \left( Q^{L}; P^{G} tan(\phi^{max}); \sqrt{\overline{S}^2 - (P^G)^2} \right)
\end{equation}


We prove analytically that LLMA provides equal or lower active power losses than the ``no-action'' strategy. Due to space limitations in this paper, the conducted theoretical proof can be found in Section II of Ref. \cite{arxiv_external}.

\subsection{Local Flow Measuring Algorithm (LFMA)}
Next, we introduce a more advanced local algorithm, which measures the incoming flows; we call it local flow measuring algorithm (LFMA). LFMA consists of four steps, and we denote the resulting variables of steps 2-4 by ${\mathcal{H}}$, ${\mathcal{I}}$, ${\mathcal{F}}$ superscripts, respectively. 
\begin{algorithm}
\caption*{\textbf{Algorithm 2: Local Flow Measuring Algorithm (LFMA)}} \label{alg:LFM}
\textbf{Step 1.} As we consider a model-free approach, branch nodes do not know in which direction is a slack bus. They determine an upstream branch, i.e. a line towards a slack bus, by selecting a branch with the biggest flow during the ``no-action'' strategy. \\
\textbf{Step 2.} All inverters follow the same procedure as during LLMA, and the reactive power generation $Q^{G,{\mathcal{H}}}$ after this step is defined by (\ref{eq:LLMA}). \\
Steps 3-4 are performed only on branch nodes, while leaf nodes do not change their own generation setpoints further.
\textbf{Step 3.} Inverters increase their own reactive generation by the value of upstream reactive flow $Q^{\mathcal{H}}_{up}$, while still satisfying the limits (\ref{eq:PF})-(\ref{eq:Slim}). The generation setpoint $Q^{G,{\mathcal{I}}}$ after step 3 is:\\
\begin{equation}\label{Alg:StepII}
Q^{G,{\mathcal{I}}} = \min \left( Q^{L} + Q^{\mathcal{H}}_{up}; P^{G} tan(\phi^{max}); \sqrt{\overline{S}^2 - (P^G)^2} \right)
\end{equation}
\textbf{Step 4.} After Step 3, reactive power flows now denoted by $Q^{\mathcal{I}}_{up}$ will be different from $Q^{\mathcal{H}}_{up}$. Step 4 is performed only if $Q^{\mathcal{I}}_{up}$ has an opposite direction from $Q^{\mathcal{H}}_{up}$. In that case, we check:
\begin{algorithmic}
	\If {downstream flow $Q_{do}$ does not change direction after Step 3, 
}
		\State decrease reactive generation by the absolute value of
		\State the measured upstream flow: 
		\begin{equation}\label{Alg:Step3c}
        Q^{G,{\mathcal{F}}} = Q^{G,{\mathcal{I}}} - |Q^{\mathcal{I}}_{up}|
        \end{equation}
	\Else 
		    \State set reactive generation $Q^{G,{\mathcal{F}}}$ according to (\ref{eq:LLMA}).
	\EndIf
\end{algorithmic} 
\end{algorithm}

There are two prerequisites for the execution of LFMA Algorithm. First, for LFMA to work effectively, we need to be able to measure the reactive power line flows at all branch nodes with an inverter. Second, to provably guarantee that LFMA converges to the same or better solution than LLMA (see Section III of Ref. \cite{arxiv_external}), all inverters are expected to perform the same step at a time. To enable that, converters can access global time settings through GPS or a simple radio-signal, similar to radio-controlled clocks, which can most often achieve an accuracy down to the exact second. 
As modern inverters allow frequent and fast change of their reactive generation settings, step 1 (if applicable) can be executed between seconds $00:10$ every minute, step 2 between $15'':25''$, step 3 (if applicable) between $30'':40''$, step 4 (if applicable) between $45'':55''$. Note that we keep blank periods between intervals of the steps to ensure their timely execution by all inverters.



\subsection{Hybrid Algorithm}
Optimal power flow (OPF) algorithms can be applied to active power loss minimization as well \cite{Murzakhanov}. The original OPF problem operates with a full vector of control variables, namely active and reactive power generation, voltage magnitude and angle \cite{Murzakhanov}. Note that the active generation of PV units is determined by the solar radiance, and only a slack bus can adjust its active power injection to maintain power balance. Finally, performing an OPF requires a central coordinator and real-time communication infrastructure.

While most OPF algorithms find the global optimum for small and medium power systems, they can arrive at suboptimal solutions for systems with thousands of nodes \cite{7879340}. Moreover, practical implementation of optimal power flow algorithms in real power systems would lead to a communication burden, when thousands of inverters are exchanging information with a central coordinator. As a result, OPF methods are presented only in academic literature, but not in real systems. In this section, we propose a hybrid algorithm, which uses a fraction of the communication needs required for the OPF solution. 

\begin{algorithm}
\caption*{\textbf{Algorithm 3: Hybrid Algorithm}} \label{alg:hybrid}
\textbf{Step 1.} Inverters execute  LLMA or LFMA. \\
\textbf{Step 2.} A central coordinator collects information on the state variables over the whole system. \\
\textbf{Step 3.} For each inverter, the central coordinator computes the remaining reactive power reserve:
\begin{equation}\label{Alg:Qres}
Q_{res} = \overline{Q}^G - Q^G
\end{equation}
\textbf{Step 4.} The central coordinator ranks inverters, depending on how much reactive power reserve $Q_{res}$ they have left. \\
\textbf{Step 5.} The central coordinator computes OPF with $N$ inverters with the highest $Q_{res}$ as control variables. 

Note that $N$ is a subset of inverters utilized during conventional OPF. The value of $N$ is defined by a central coordinator by taking into consideration the system size, communication infrastructure, and available computation power.
\end{algorithm}

The idea behind using (\ref{Alg:Qres}) as a criterion for selecting centrally controlled inverters is the following. Higher $Q_{res}$ provides a broader range of possible setpoints for an inverter. As a result, the probability to select setpoints leading to a more optimal solution is higher. Note that as PV output and load consumption varies with time, the inverters selected by a central coordinator may change too.

While simple, Algorithm 3 resolves the computation burden issue, which is typical for OPF solutions in large systems: it involves a smaller number of control variables and, thus, it results in a simpler optimization problem and faster calculation. On top of that, we gain additional computation speed through the decreased communication burden, as we only need to communicate the computed setpoints to a fraction of inverters.

\subsection{Impact on Nodal Voltages of the Proposed Algorithms}
The proposed LLMA and LFMA reduce losses by adjusting the reactive power generation setpoints so that the reactive power flows from a slack node decrease compared to the ``no-action'' strategy. This also means that they result to lower voltage drops between neighboring buses. As a result, the nodal voltages are expected to come closer to the reference bus voltage, which in most distribution is set by the slack bus. The slack bus is also often equipped with a voltage regulator or load-tap-changer to adjust voltage to the desired level. Compared to the ``no-action'' strategy, voltages that were lower than the slack bus voltage will increase, while voltages that were higher than the slack bus voltage will decrease. Assuming that during the ``no-action'' strategy all nodal voltages were within limits, applying the LLMA or the LFMA will maintain the voltages within the same limits, or even move them closer together and further away from the bounds. This is indeed what we observe in all the simulations we carried out in Section VI, for various cases, as also shown in Table VI.


\subsection{Applicability of the Algorithms in Unbalanced Systems}
Power distribution systems can be modeled either as three-phase systems or by their single-phase equivalents. However, even in more inverter-specific studies, it is a common practice to model a single three-phase inverter as three single-phase ones \cite{8442493}. This results in higher flexibility and reduces complexity while still representing equally well the inverter capabilities for steady-state studies. As a result, in a three-phase unbalanced system, our LLMA and LFMA algorithms for the control of the inverters will apply in exactly the same way, in each phase separately, with all operation principles remaining the same. That is why further we provide numerical tests for a single-phase equivalent of the considered systems.

\section{Numerical results} \label{sec:NumRes}
In this section, we provide numerical results for the 5-bus system in Fig.~\ref{fig:Math_System}, the IEEE 141-bus radial network, a part of the Danish distribution system, and the meshed IEEE 30-bus system. The code to reproduce the reported results is available online \cite{Ilgiz_code}. We show that LLMA provides lower active power losses than the ``no-action'' strategy and that LFMA obtains lower active power losses than LLMA. We demonstrate that these findings hold under varying solar generation and consumption scenarios. Moreover, we illustrate several examples proving that LLMA and LFMA are robust to topology reconfigurations. Finally, we show that hybrid LFMA outperforms centralized OPF in terms of higher optimization capability and lower computation and communication burden.

Note that in all numerical simulations voltage magnitudes at all nodes are within operational limits $[0.90; 1.10]$ p.u.

\subsection{Explanation of the Algorithms on the 5-bus System}
In this section, we illustrate the performance of  the ``no-action'' strategy,  the local load and local flow measurement algorithms on the 5-bus system in Fig.~\ref{fig:Math_System}. For compactness, we display only reactive generation setpoints as they are the only control variables in LLMA and LFMA, while active generation is defined by time-varying PV output. Similarly, only reactive power flows and reactive loads are displayed in lines and nodes, accordingly. We provide values of reactive power generation limits and reactive loads in Table~\ref{tab:5bus}. Note that the values provided in Table~\ref{tab:5bus} and Figs.~\ref{fig:5bus_noAct}-\ref{fig:5bus_LFMA_step3} are given in kVAr.

\begin{table}[t]
\centering
\caption{Data of the 5-bus system.}
\label{tab:5bus}
\begin{tabular}{ccc}
Node & $\overline{Q}^G$, (kVAr) & $Q^L$, (kVAr) \\  \hline
2    &   9.00  &  7.00   \\
3    &   6.00  &  4.00   \\
4    &   2.40  &  3.00   \\
5    &   0.00  &  1.00  
\end{tabular}
\end{table}

\begin{figure}[H]
    \centering
    \includegraphics[width=0.9\linewidth]{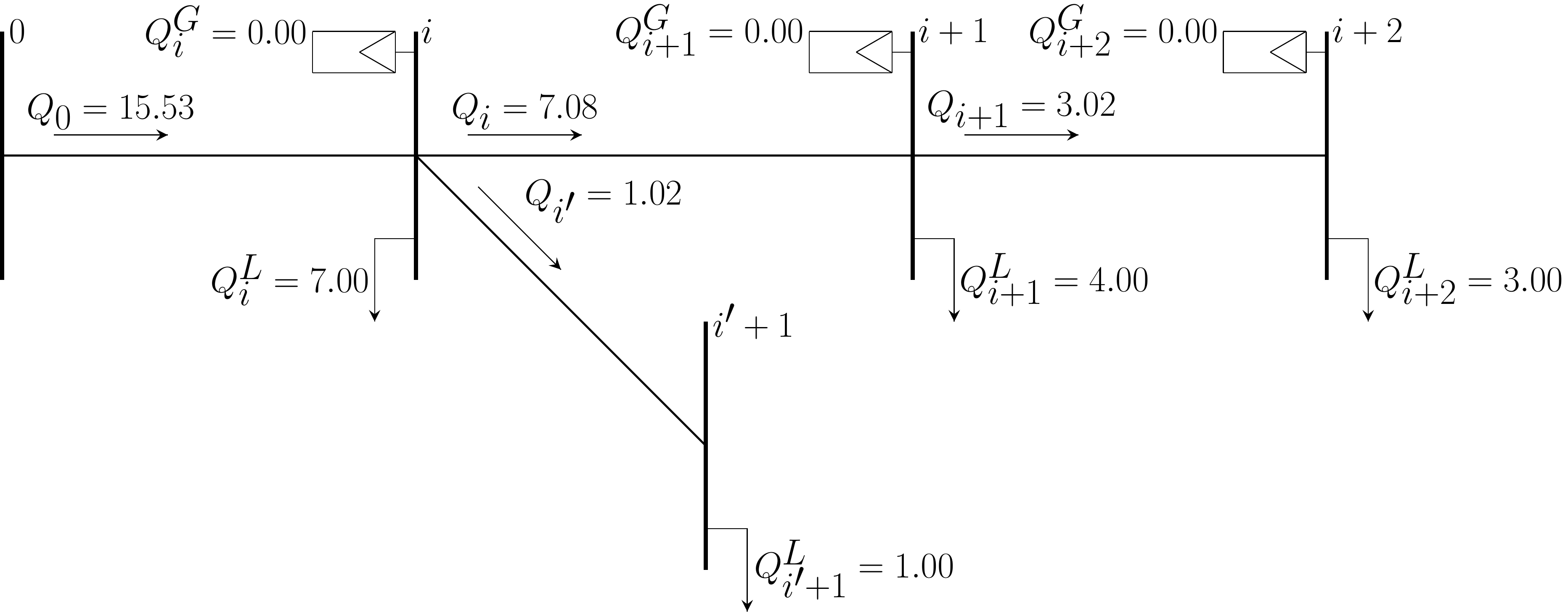}
    \caption{The ``no-action'' strategy on the 5-bus system.}
    \label{fig:5bus_noAct}
\end{figure}

\subsubsection{Application of the ``no-action'' strategy [see~Fig.~\ref{fig:5bus_noAct}]} 
local reactive generation is set to zero, and all the reactive demand is supplied by the slack node $0$. Note the high values of reactive power flows.

\begin{figure}[H]
    \centering
    \includegraphics[width=0.9\linewidth]{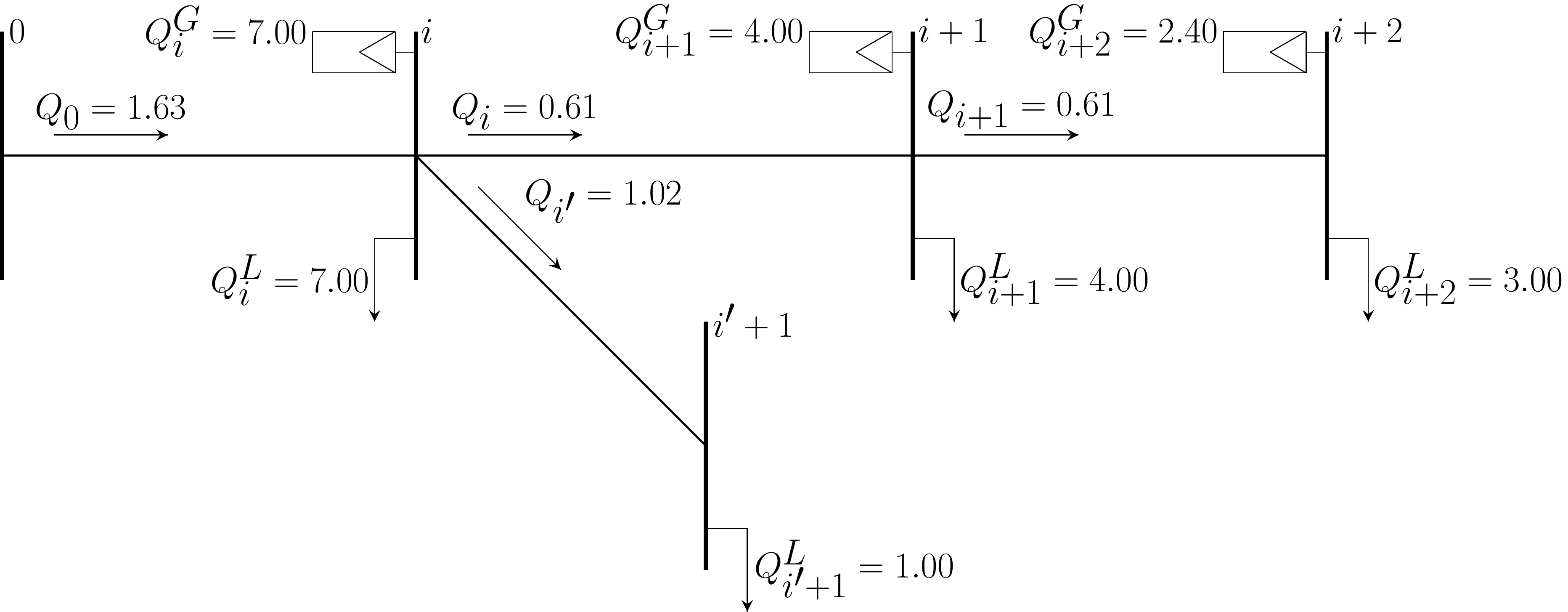}
    \caption{LLMA on the 5-bus system.}
    \label{fig:5bus_LLMA}
\end{figure}

\subsubsection{Application of the local load measuring algorithm (LLMA) [see~Fig.~\ref{fig:5bus_LLMA}]} 
only nodes $i$ and $i+1$ have sufficient reactive power capacities to cover their own loads. Part of load in node $i+2$ is covered by the slack node, and the full load in bus $i'+1$ is covered by the slack node. Note that application of LLMA leads to lower power flows; thus, to lower power losses. For example, $Q_i=0.61$ with LLMA, while $Q_i=7.08$ in the ``no-action'' strategy.

\subsubsection{Application of the local flow measuring algorithm (LFMA)} 
LFMA consists of four steps, see Algorithm 2. 

In step 1 of LFMA, each inverter determines the upstream branch by selecting a line with the biggest power flow. In Fig.~\ref{fig:5bus_noAct}, we see that inverters $i$ and $i+1$ would select lines $0$ and $i$, accordingly. 

\begin{figure}[H]
    \centering
    \includegraphics[width=0.9\linewidth]{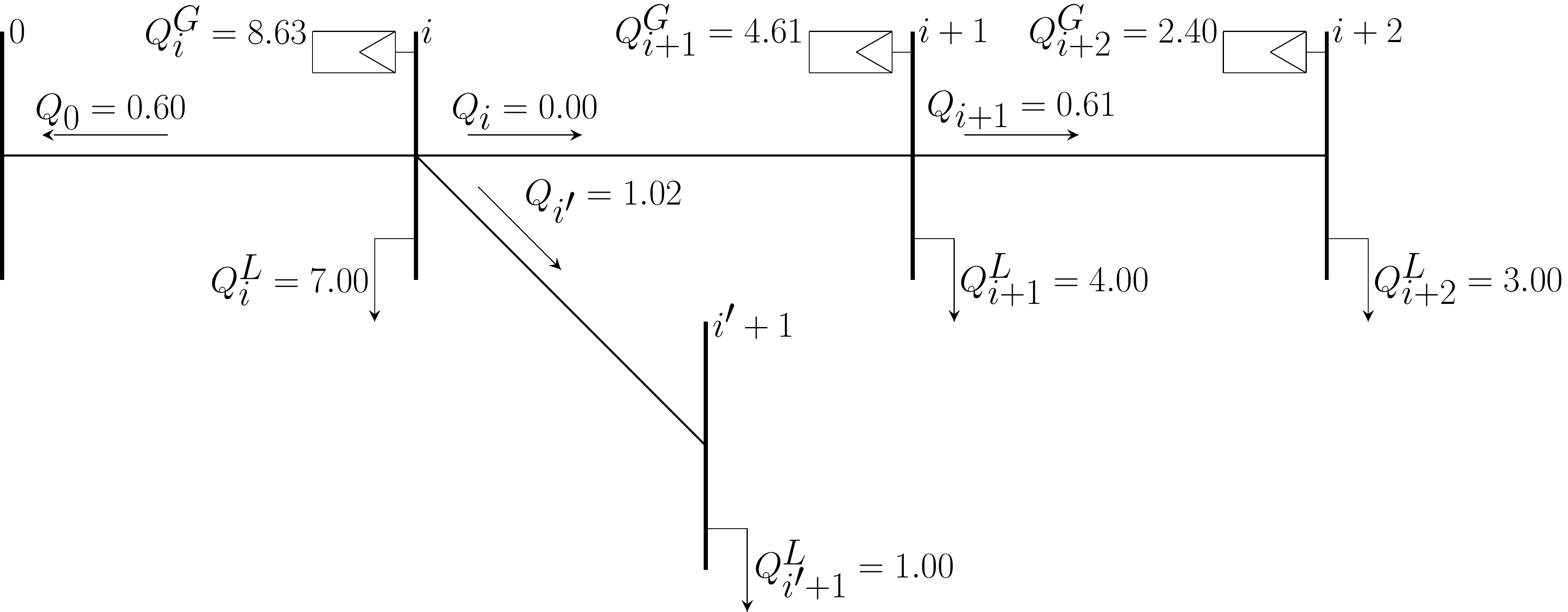}
    \caption{Step 3 of LFMA on the 5-bus system.}
    \label{fig:5bus_LFMA_step2}
\end{figure}

Step 2 of LFMA is equivalent to LLMA, so it is shown in Fig.~\ref{fig:5bus_LLMA}. Application of step 3 is shown in Fig.~\ref{fig:5bus_LFMA_step2}. 

\begin{figure}[H]
    \centering
    \includegraphics[width=0.9\linewidth]{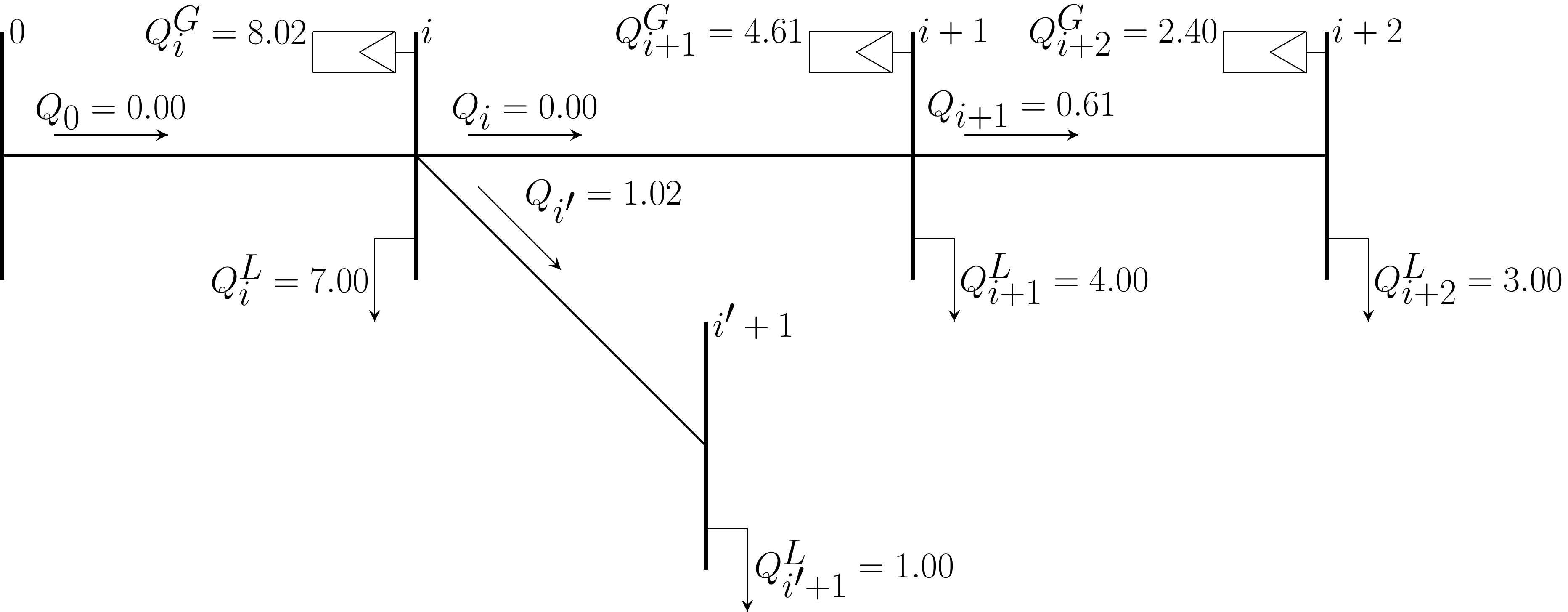}
    \caption{Step 4 of LFMA on the 5-bus system.}
    \label{fig:5bus_LFMA_step3}
\end{figure}

Note that the upstream flow of bus $i$ changes its own direction between steps 2 and 3. Thus, step 4 is performed only by an inverter in bus $i$, and its application is shown in Fig.~\ref{fig:5bus_LFMA_step3}. Note that power flows in lines $0$ and $i+1$ decrease even further compared to LLMA in Fig.~\ref{fig:5bus_LLMA}; thus, lower power losses in a system are obtained.

\subsection{Simulations for the IEEE 141-bus System}
In this section, we implement Algorithms 1-3 and compare them with  the ``no-action'' strategy and  the centralized OPF solutions in the IEEE 141-bus system. The system has 140 branches and 84 loads with a total nominal demand of 11.94 MW and 7.40 MVAr. The original IEEE 141-bus system does not contain any distributed generation units \cite{Matpower}. In order to measure the performance of our algorithms without being dependent on the specific placement of the PV inverters, we generate 1'000 random instances in each of which we randomly place 30 PVs in the 141-bus system, and apply our algorithms. We do the same for 60 PVs and 80 PVs. To objectively measure how our algorithms perform with `more distributed' or `less distributed' generation (i.e. many and small or few and larger DER) we maintain the total installed capacity of PVs the same across the cases of 30, 60, 80 PVs (and instead adjust uniformly the installed capacity of every single PV inverter). We report the mean value and standard deviation for the losses in each case after the 1'000 random placements. As there is only one power loss value for the original system (i.e. since it did not contain any PVs), its value is given as a mean. In addition to the local load and local flow algorithms, we implement Algorithm 3 after performing LLMA or LFMA, and we refer to it as hybrid LLMA and hybrid LFMA, respectively. Note that the number of centrally controlled inverters may vary for hybrid LLMA and hybrid LFMA. In all implemented algorithms, the voltage limits $[0.90; 1.10]$ p.u. are satisfied. The results are provided in Table~\ref{tab:141bus}.  

There are several observations from Table \ref{tab:141bus}. First, comparing original and distributed-30 systems, we conclude that adding 30 PVs can decrease active power losses by more than $67\%$. Second, from the comparison of all distributed systems, it follows that a greater number of PVs leads to smaller values of the mean and standard deviation of active power losses for all approaches. This is because a larger number of PVs leads to a shorter path between generation and consumption, and therefore lower losses. Third, comparing different algorithms within each distributed system type, we see that LLMA always provides a lower mean of power losses than the ``no-action'' strategy, and LFMA always obtains lower mean of power losses than LLMA. Also, we see that hybrid LLMA and hybrid LFMA provide the same results as  the centralized OPF, but require a fewer number of centrally controlled inverters. This is because of the communication-free and model-free algorithms we propose (LLMA and LFMA) that act in step 1 of Algorithm 3. Notably, hybrid LFMA requires fewer centrally controlled inverters than hybrid LLMA; at the expense, though, of the need to additionally measure reactive power flows (hybrid LLMA only needs to measure the local reactive power demand). 

\begin{figure}[htp]

\subfloat[30 PVs]{%
  \includegraphics[clip,width=\columnwidth]{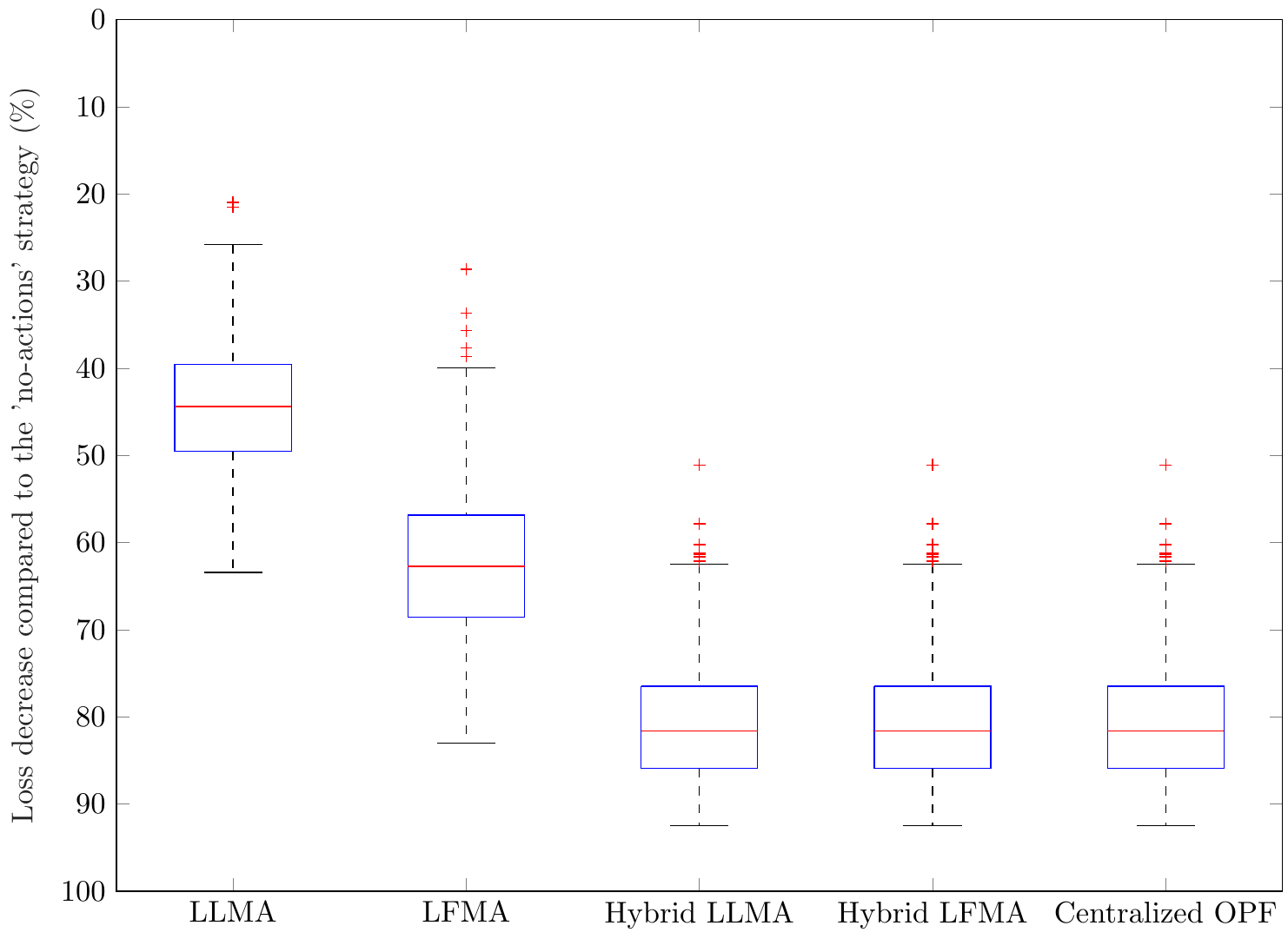}%
}

\subfloat[60 PVs]{%
  \includegraphics[clip,width=\columnwidth]{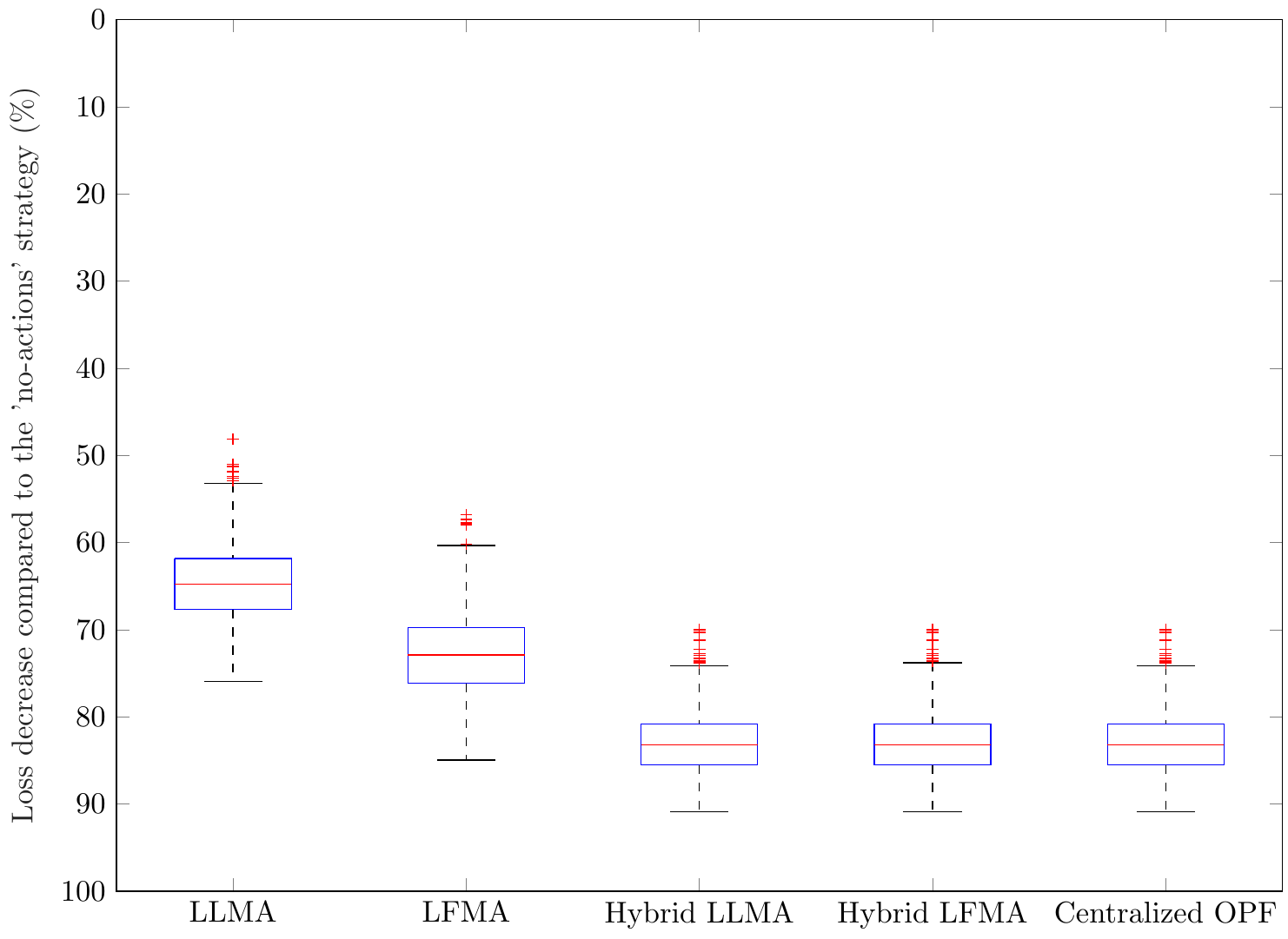}%
}

\subfloat[80 PVs]{%
  \includegraphics[clip,width=\columnwidth]{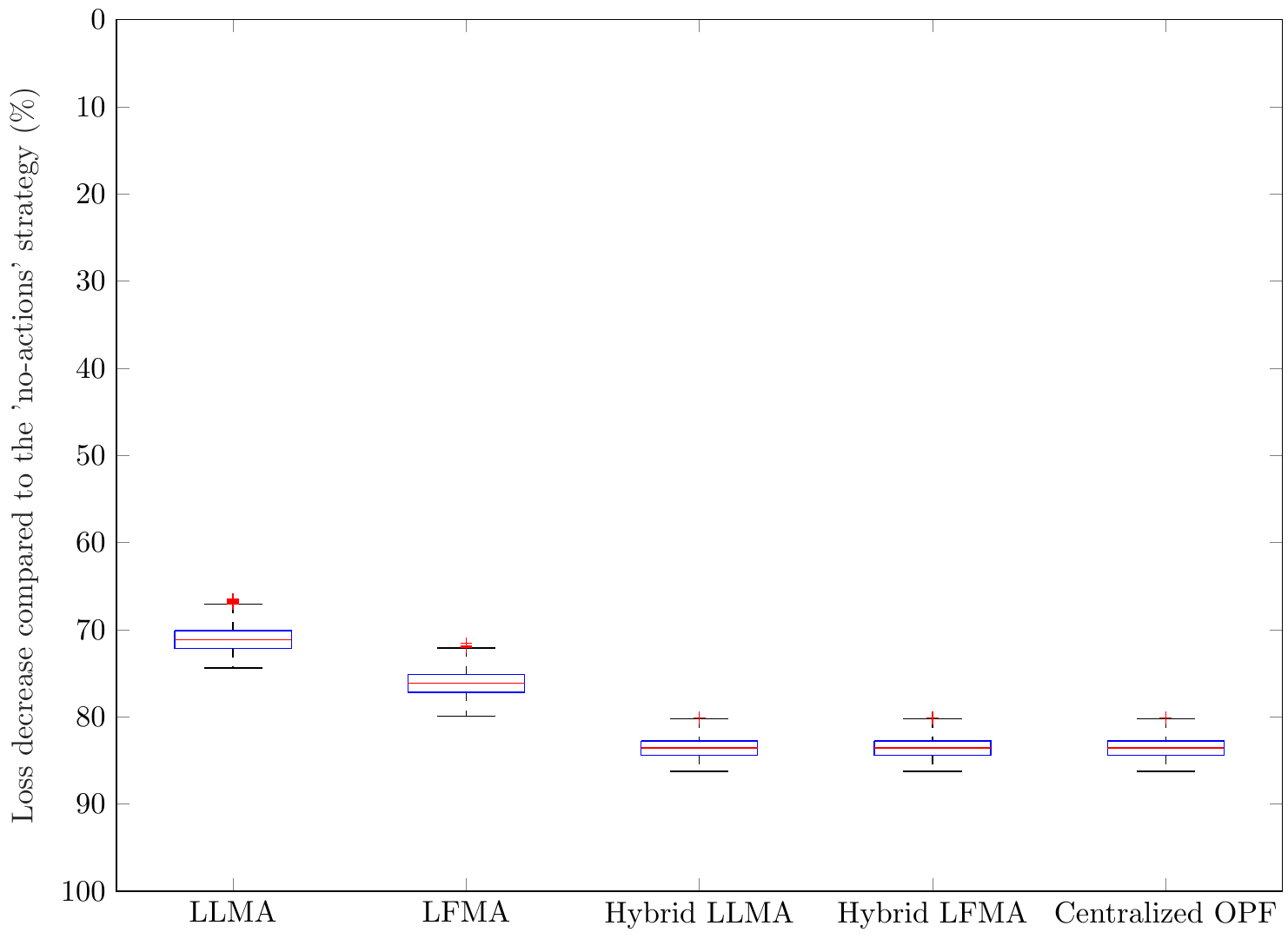}%
}

\caption{Power loss decrease (in \%) in percent by the communication-based and communication-free approaches compared to the ``no-action'' strategy for 1'000 times random placement of PVs in the IEEE 141-bus system.}
\label{fig:boxplot}
\end{figure}

The power loss decrease in percent by the communication-based and communication-free approaches compared to the ``no-action'' strategy is displayed on box plots in Fig.~\ref{fig:boxplot}, and we conclude the following. First, LLMA and LFMA obtain up to $76\%$ and $85\%$ power loss decrease, respectively. Second, both hybrid algorithms achieve to decrease losses up to $94\%$, which is the same as for centralized OPF. Third, a higher number of PVs leads to more narrow distribution for each of the depicted algorithms in Fig.~\ref{fig:boxplot}. We explain it by the fact that in 1'000 simulations, PVs are randomly placed in the system. As a result, there are many more variations of placing 30 identical PVs in the 141-bus system, than placing 80 PVs.

Distribution grids originally have a loopy graph, while they are operated in radial topology. The topology reconfiguration is obtained from the original graph by opening switches on some lines and closing on others. One reason behind topology reconfiguration is the maintenance operations of power lines. It is obvious that we want our distribution grid algorithms to be robust (ideally agnostic) to any topology changes. To assess if our algorithms maintain the same performance under topology changes, we consider three cases where one line is switched off, and another is switched on, to model the aforementioned scenarios of line faults. We conduct the simulations for the network with 1'000 random locations of 30 PVs, and we present the results in Table \ref{tab:141bus_topCh}. Comparing Tables \ref{tab:141bus} and \ref{tab:141bus_topCh}, we can make exactly the same observations for Table \ref{tab:141bus_topCh} as we did for Table \ref{tab:141bus}: LFMA performs better than LLMA and the hybrid algorithms perform better than the purely local ones. 
We conclude that the aforementioned observations on the local measuring and hybrid algorithms in Table \ref{tab:141bus} hold in Table \ref{tab:141bus_topCh} as well. 
What is important though is, that LLMA and LFMA manage again to considerably reduce the losses compared with the ``no-action'' strategy, while being at the same time completely robust to any topology changes, as they operate with only local information. This can be of their strengths when it comes to their possible implementation in real distribution grids.

\begin{table*}[t]
\centering
\begin{center}
\caption{Comparison of the communication-based and communication-free algorithms in the IEEE 141-bus system for different total number of installed PVs; in each of the three cases, we report the mean and standard deviation after 1'000 random PV placements.}
\label{tab:141bus}
\begin{tabular}{p{0.1\textwidth}<{\centering}p{0.15\textwidth}<{\centering}p{0.15\textwidth}<{\centering}p{0.08\textwidth}<{\centering\arraybackslash}p{0.08\textwidth}<{\centering}p{0.15\textwidth}<{\centering}}
\hline
System type & Number of PVs & Algorithm & \multicolumn{2}{c}{Active power losses, (kW)} & Average number of centrally controlled inverters \\ \cline{4-5} 
                  &                   &                   &  mean & std  &  \\ \hline
Original & 0              &  -  &  629.06 &  ~~~~~- & ~0 \\ \hline
\multirow{6}{*}{Distributed-30} & \multirow{6}{*}{30} & ``No-action'' strategy &  206.88 & 18.45 & ~0 \\ 
                  &                   & LLMA & 116.15 & 24.79 & ~0\\ 
                  &                   & LFMA & ~79.58 & 25.61 & ~0 \\ 
                  &                   & Hybrid LLMA & ~41.19 & 18.43 & 28 \\ 
                  &                   & Hybrid LFMA & ~41.19 & 18.43 & 21 \\                   
                  &                   &  Centralized OPF & ~41.19 & 18.43 & 30 \\ \hline
\multirow{6}{*}{Distributed-60} & \multirow{6}{*}{60} & ``No-action'' strategy & 200.78 & ~8.71 & ~0 \\ 
                  &                   & LLMA & ~71.45 & 11.70 & ~0 \\ 
                  &                   & LFMA & ~55.04 & 11.73 & ~0 \\ 
                  &                   & Hybrid LLMA & ~34.49 & ~8.61 & 46 \\ 
                  &                   & Hybrid LFMA & ~34.49 & ~8.61 & 35 \\  
                  &                   & Centralized OPF & ~34.49 & ~8.61 & 60 \\ \hline                   
\multirow{6}{*}{Distributed-80} & \multirow{6}{*}{80} & ``No-action'' strategy  & 199.30 & ~3.03 & ~0 \\  
                  &                   & LLMA & ~57.74 & ~3.92 & ~0 \\ 
                  &                   & LFMA & ~47.69 & ~3.78 & ~0 \\ 
                  &                   & Hybrid LLMA & ~32.85 & ~3.00 & 59 \\ 
                  &                   & Hybrid LFMA & ~32.85 & ~3.00 & 44 \\  
                  &                   & Centralized OPF &  ~32.85 & ~3.00 & 80 \\ \hline                
\end{tabular}
\end{center}
\end{table*}

\begin{table*}[t]
\centering
\begin{center}
\caption{Comparison of the communication-based and communication-free algorithms in the IEEE 141-bus system with 1'000 times random placement of 30 PVs under topology reconfiguration cases.}
\label{tab:141bus_topCh}
\begin{tabular}{p{0.15\textwidth}<{\centering}p{0.15\textwidth}<{\centering}p{0.15\textwidth}<{\centering}p{0.08\textwidth}<{\centering\arraybackslash}p{0.08\textwidth}<{\centering}p{0.15\textwidth}<{\centering}}
\hline
Switched-off line & Switched-on line & Algorithm & \multicolumn{2}{c}{Active power losses, (kW)} & Average number of centrally controlled inverters \\\cline{4-5} 
                  &                   &                   &  mean & std \\ \hline
\multirow{6}{*}{5-6} & \multirow{6}{*}{7-34} & ``No-action'' strategy & 150.14 & 18.86 & ~0 \\ 
                  &                   & LLMA & ~88.16 & 24.13  & ~0 \\ 
                  &                   & LFMA & ~64.09 & 24.63  & ~0 \\ 
                  &                   & Hybrid LLMA & ~36.64 & 18.71 & 28 \\ 
                  &                   & Hybrid LFMA & ~36.64 & 18.71 & 21 \\    
                  &                   &  Centralized OPF & ~36.64 & 18.71  & 30 \\ \hline
\multirow{6}{*}{15-118} & \multirow{6}{*}{17-130} & ``No-action'' strategy & 208.14 & 19.01 & ~0 \\ 
                  &                   & LLMA & 117.04 & 25.25 & ~0 \\ 
                  &                   & LFMA  & ~80.22 & 26.32 & ~0 \\  
                  &                   & Hybrid LLMA & ~41.77 & 18.94 & 28 \\ 
                  &                   & Hybrid LFMA & ~41.77 & 18.94 & 21 \\    
                  &                   & Centralized OPF & ~41.77 & 18.94 & 30 \\ \hline            
\multirow{6}{*}{76-78} & \multirow{6}{*}{45-82} & ``No-action'' strategy  & 208.16 & 18.62 & ~0 \\  
                  &                   & LLMA & 117.13 & 25.07 & ~0 \\ 
                  &                   & LFMA  & ~80.94 & 25.80 & ~0 \\  
                  &                   & Hybrid LLMA & ~41.86 & 18.60 & 28 \\ 
                  &                   & Hybrid LFMA & ~41.86 & 18.60 & 21 \\    
                  &                   & Centralized OPF & ~41.86 & 18.60 & 30 \\ \hline              
\end{tabular}
\end{center}
\end{table*}

\begin{table*}[t]
\centering
\begin{center}
\caption{Comparison of the communication-based and communication-free algorithms in the 161-bus Danish distribution system in the full year 2019.}
\label{tab:Akirkeby}
\begin{tabular}{p{0.15\textwidth}<{\centering}p{0.1\textwidth}<{\centering\arraybackslash}p{0.1\textwidth}<{\centering}p{0.15\textwidth}<{\centering}p{0.13\textwidth}<{\centering}p{0.15\textwidth}<{\centering}}
\hline
Algorithm & \multicolumn{2}{c}{Electricity losses} & Savings w.r.t.  the ``no-action'' strategy, & Number of infeasible cases & Average number of centrally controlled \\ \cline{2-3} 
                   & MWh/year & \euro/year & (\euro/year) & out of 99 704 & inverters \\ \hline 
``No-action'' strategy &  18 194.76 & 4 690 427.94 & ~~~~~~- & ~~~~~~0 & ~0 \\ 
LLMA & 18 007.64 & 4 642 189.55 & ~48 238.39 & ~~~~~~0 & ~0 \\ 
LFMA & 17 958.47 & 4 629 515.27 & ~60 912.66  & ~~~~~~0 & ~0 \\ 
Hybrid LLMA & 16 785.93  & 4 327 245.64 & 363 182.30 & 55 364 & 11  \\ 
Hybrid LFMA & 16 785.76 & 4 327 203.57 & 363 224.36  & 55 364 & ~8  \\                   
Centralized OPF & 16 785.78 & 4 327 207.42 & 363 220.51  & 55 364 & 36  \\ \hline
\end{tabular}
\end{center}
\end{table*}

\begin{table*}[t]
\centering
\begin{center}
\caption{Comparison of the communication-based and communication-free algorithms in the meshed IEEE 30-bus system.}
\label{tab:meshedSys}
\begin{tabular}{p{0.1\textwidth}<{\centering}p{0.15\textwidth}<{\centering}p{0.15\textwidth}<{\centering}p{0.15\textwidth}<{\centering}p{0.15\textwidth}<{\centering}}
\hline
System type & Number of DERs & Algorithm & Active power losses, (kW) & Number of centrally controlled inverters \\ \hline
\multirow{6}{*}{IEEE 30-bus system} & \multirow{6}{*}{5} & ``No-action'' strategy & 5 048.61 & 0 \\ 
                  &                   & LLMA & 2 625.97 & 0 \\ 
                  &                   & LFMA & 2 494.78 & 0 \\ 
                  &                   & Hybrid LLMA & 2 432.70 & 5 \\ 
                  &                   & Hybrid LFMA & 2 432.70 & 5 \\                   
                  &                   & Centralized OPF & 2 432.70 & 5 \\ \hline                   
\end{tabular}
\end{center}
\end{table*}

\begin{table*}[t]
\centering
\begin{center}
\caption{Comparison of minimum and maximum voltage magnitudes over all buses in the IEEE 30-bus system during operation with and without LTC control.}
\label{tab:ltc}
\begin{tabular}{p{0.15\textwidth}<{\centering}p{0.08\textwidth}<{\centering\arraybackslash}p{0.08\textwidth}<{\centering}p{0.01\textwidth}<{\centering}p{0.08\textwidth}<{\centering\arraybackslash}p{0.08\textwidth}<{\centering}}
\hline
Algorithm & \multicolumn{2}{c}{Without LTC control} & & \multicolumn{2}{c}{With LTC control} \\ \cline{2-3} \cline{5-6} 
                   &  $V_{min}$ &  $V_{max}$  & &  $V_{min}$ & $V_{max}$  \\ \hline
``No-action'' strategy  & 1.06 & 1.13 & & 1.06 & 1.09 \\  
LLMA & 1.06 & 1.13 & & 1.06 & 1.09 \\ 
LFMA & 1.06 & 1.13 & & 1.06 & 1.08 \\ 
Hybrid LLMA & 1.06 & 1.13 & & 1.06 & 1.09 \\ 
Hybrid LFMA & 1.06 & 1.13 & & 1.06 & 1.09 \\  
Centralized OPF &  1.06 & 1.13 & & 1.06 & 1.08 \\ \hline                
\end{tabular}
\end{center}
\end{table*}

\subsection{Simulations for a Part of the Danish Distribution System}
In this section, we implement the  communication-based and communication-free algorithms on a part of the Danish distribution system. The considered part of the radial distribution network has 161 buses and 160 branches. There are 36 distributed energy sources and 97 consumer nodes with a nominal load of 8.18 MW and 4.22 MVAr. We utilize five-minute-based solar generation and active power demand data for the full year 2019 provided by SYSLAB \cite{syslab}. Based on the solar generation and power demand data, we compute the estimated cost savings from reducing the electricity losses in terms of MWh and euros. In addition, we report a number of infeasible simulations for each algorithm type. The results are given in Table~\ref{tab:Akirkeby}. We conclude that implementation of  the local load and local flow measuring algorithms saves around 48 and 61 thousand euros per year compared to the ``no-action'' strategy, respectively. At the same time, with a limited need for communication compared to the centralized algorithm, the developed hybrid algorithms show a much higher potential for reducing losses and may save around 363 thousand euros per year. As a matter of fact, to achieve the same reduction in losses, hybrid LLMA and hybrid LFMA require 3-5 times fewer centrally controlled inverters than the centralized OPF. 

Note that the local load and local flow algorithms are robust and provide solutions in all 99704 time steps. On the contrary, the algorithms that utilize optimal power flow calculations, namely hybrid LLMA, hybrid LFMA, and the centralized OPF, experience computational issues in more than $55\%$ of cases. During these cases, OPF cannot provide optimal or even feasible setpoints. When this happens, we choose to consider the last previously known setpoints of the local algorithms in the optimization.
More specifically, if hybrid LLMA or hybrid LFMA fail to converge, then the setpoints of LLMA or LFMA for the previous time step are used, accordingly. In contrast, if the centralized OPF fails, then the ``no-action'' strategy is performed, as we assume that communication-free algorithms are not established in that case. As we see in Table~\ref{tab:Akirkeby}, the combination of a local communication-free algorithm with the central control of a very limited number of inverters (e.g. hybrid LFMA requires only 22\% of the inverters used in centralized OPF) can lead to results that are even better than having a full communication and control of all PVs in the system.


\subsection{Simulations for the Meshed IEEE 30-bus System}
Proposed algorithms can be operated in any type of power distribution system, including meshed systems. Although we have derived analytical proofs for the performance of LLMA and LFMA only for radial systems, since we use the DistFlow model, in practice, LLMA and LFMA work equally well for the meshed systems as well. To demonstrate the applicability of the proposed algorithms for realistic meshed systems, we implement them in the meshed IEEE 30-bus system \cite{Matpower}. Note that this system has switched capacitors, transformers, and load tap changers. As shown in Table~\ref{tab:meshedSys}, all the algorithms, including LLMA and LFMA, perform exactly as expected in the case of the meshed system with various types of equipment, and similar to the radial cases.

We consider two additional scenarios of solar power generation and load from real SYSLAB data to demonstrate the performance of load tap changers (LTCs). We compute the minimum and maximum voltage magnitudes across all buses for different algorithms and present the results in Table~\ref{tab:ltc}. The tested IEEE 30-bus system has four transformers, that are equipped with LTCs. As we see from Table~\ref{tab:ltc}, without LTC control the maximum voltages across all methods exceed the allowed limit of $1.1$ p.u. Deploying LTCs allows to bring voltages across all methods in the permitted range of $[0.90; 1.10]$ p.u.. We see that in both cases, with and without the LTCs, our algorithms perform as expected.

\subsection{Voltage Ranges in the Proposed Algorithms}
In all conducted simulations for various cases our proposed algorithms kept voltages within $[0.90; 1.10]$ p.u. as long as voltages during the ``no-action'' strategy were also in $[0.90; 1.10]$ p.u. range. These results, as shown in Table~VI, numerically confirm the discussion based on the underlying theory we carried out in Section V.D: the proposed local and hybrid algorithms keep the same or higher voltage for $V_{min}$, and have the same or lower voltage for $V_{max}$ compared to the ``no-action'' strategy.

\section{Discussion}\label{sec:disc}
The operation of the current distribution grids is not optimized by the centralized OPF due to several problems.

Current and future distribution grids will experience the connections of millions of inverter-connected resources (solar PVs, batteries, electric vehicles, heat pumps, etc.) and much higher flows. To avoid excessive grid investments, non-wire solutions that rely on actively controlling the available inverters will become necessary. The approaches we propose in this paper reduce the grid losses -- and, thus, the system loading -- through communication-free and model-free algorithms which only adjust the reactive power setpoints using the active front-end control of grid-connected inverters. These have three distinct benefits. First, by not affecting the active power injections of the inverters, our algorithms do not interfere at all with any financial transactions between the consumers and the grid operator (e.g. peer-to-peer markets, balancing, demand response, etc.).

Second, they are highly scalable, requiring a much lower communication and computational burden compared to a centralized algorithm. Our hybrid LLMA and hybrid LFMA allow the operator to determine the number of centrally controlled inverters. As shown in Tables~II, IV, and VII, the hybrid methods require a fewer number of centrally controlled inverters while achieving the same or very close results in loss minimization as the centralized OPF. We consider that this property has high practical value for the distribution grids of sizes that exceed tens of thousands of buses.

Third, the proposed LLMA and LFMA approaches counter to a certain extent issues related to poor observability in distribution networks, due to inexistent communication with the system operator or out-of-date models in the operator's database. This problem greatly reduces the capability of the operator to provide optimal setpoints for these unobserved parts of the grid. Our LLMA and LFMA methods resolve this problem since they do not require any communication; they only need local information for computing optimal setpoints of inverters. In fact, LLMA and LFMA can be seen as special cases of hybrid LLMA and hybrid LFMA, respectively, when the communication is permanently or temporarily unavailable.

Finally, the proposed algorithms have plug'n'play capabilities and are topology agnostic: being applicable in real distribution systems (Section~VI.C) under topology reconfiguration (Section~VI.B), in systems with meshed topology and equipped with various discrete devices (Section~VI.D), is what makes them valuable for practical use.

\section{Conclusions and future work}\label{sec:conclus}
In this paper, we present four algorithms for optimizing modern distribution grids that undergo massive penetration of DERs. With millions of converter-interfaced devices connected to the grid, the sheer computation and communication requirements to centrally control all devices render optimization algorithms incapable to achieve that in real-time. Focusing on the problem of loss minimization in distribution grids, this paper proposes two communication-free and model-free algorithms that can act locally within each inverter. We analytically prove that both algorithms reduce grid losses by only controlling the reactive power setpoint of the inverters, while requiring no prior information about the network, no communication, and based only on local measurements. As we show, both algorithms are topology and network agnostic, offering plug'n'play capabilities. Going a step further, we combine the two proposed algorithms with a central optimization of a limited number of resources. We show that the hybrid approaches we propose achieve the same reduction in losses as a fully centralized algorithm, but require the central control of up to 5 times fewer resources while also offering performance guarantees in case of communication failure. We demonstrate our algorithms on the 5-bus network, the IEEE 141-bus system, the real Danish distribution system, and the meshed IEEE 30-bus system with various types of equipment.

Future work includes the development of advanced hybrid algorithms robust to incomplete or partly false topology information, and the demonstration of all algorithms in experimental facilities that include hardware-in-the-loop and real-time simulations of a real system.

\bibliographystyle{IEEEtran}
\bibliography{library}

 \end{document}